\documentclass[referee]{raa}           

\usepackage{natbib}
\usepackage{amssymb,amsmath}
\bibpunct{(}{)}{;}{a}{}{,}
\usepackage{rotating} 
\usepackage{subcaption}
\usepackage{caption}
\usepackage{graphicx}
\usepackage{adjustbox}

\usepackage[pagebackref=true]{hyperref}

\newcommand{\ha}{\hbox{H$\alpha$}}
\newcommand{\nii}{\hbox{[N\,{\sc II}]}}
\newcommand{\sii}{\hbox{[S\,{\sc II}]}}
\newcommand{\oiii}{\hbox{[O\,{\sc III}]}}

\newcommand{\continuedcaption}[1]{%
  \addtocounter{figure}{-1} 
  \caption{(Continued) #1} 
}

\begin{document}

\title{The size-velocity dispersion relationship of Galactic \hii\, regions}

\author{Lin Ma\inst{1,2,3}, Yunning Zhao$^\dagger$\inst{2,3}, Wei Zhang$^{\dagger\dagger}$\inst{2}, Youliang Feng$^*$\inst{1}, Shiming Wen\inst{2}, Shichao Han\inst{4,5}, Chaojian Wu\inst{2}, Juanjuan Ren\inst{2}, Jianjun Chen\inst{2}, Yuzhong Wu\inst{2}, Zhongrui Bai\inst{2}, Yonghui Hou\inst{3,6}, Yongheng Zhao\inst{2,3}, Hong Wu\inst{2,3}}

\email{Yunning Zhao$^\dagger$: ynzhao@bao.ac.cn, Wei Zhang$^{\dagger\dagger}$: xtwfn@bao.ac.cn, Youliang Feng$^*$: fengyouliang@utibet.edu.cn}
 
\institute{The Key Laboratory of Cosmic Rays (Xizang University), Ministry of Education, Lhasa 850000, Tibet, People's Republic of China\\
\and
National Astronomical Observatories,  Chinese Academy of Sciences, Beijing 100101, People's Republic of China\\
\and
School of Astronomy and Space Science, University of Chinese Academy of Sciences, Beijing 100049, People's Republic of China\\
\and
Cosmic Dawn Center (DAWN), Copenhagen, Denmark\\
\and 
Niels Bohr Institute, University of Copenhagen, Jagtvej 128, 2200 Copenhagen N, Denmark\\
\and
Nanjing Institute of Astronomical Optics, \& Technology, National Astronomical Observatories, Chinese Academy of Sciences, Nanjing 210042, People's Republic of China\\
\vs \no
   {\small Received 20XX Month Day; accepted 20XX Month Day}
}

\abstract{
The size-velocity dispersion ($\sigma$) relation, while well established for giant \hii\, regions, remains uncertain for their smaller counterparts (physical radii R $<$ 20 pc). Thanks to the LAMOST MRS-N dataset's large sky coverage and high spatial/spectral resolution, we examined this relationship using 10 isolated Galactic \hii\, regions with  R $<$ 20 pc. Our results reveal two key findings: (1) these small-size \hii\, regions remarkably follow the same size-$\sigma$ relation as giant \hii\, regions, suggesting this correlation could serve as a novel distance indicator for Galactic \hii\, regions; and (2) we find distinct dynamical behaviors between younger and older \hii\, regions. 
Specifically, in younger ($<$ 0.5 Myr), ionization-bounded \hii\, regions, the velocity dispersion shows no correlation with expansion velocity, indicating that turbulence is driven primarily by stellar winds and ionization processes. In contrast, in older ($>$ 0.5 Myr), matter-bounded \hii\, regions, a clear correlation emerges, implying that expansion-driven processes begin to play a significant role in generating turbulence. We therefore propose an evolutionary transition in the primary turbulence mechanisms—from being dominated by stellar winds and radiation to being increasingly influenced by expansion-driven dynamics—during the evolution of HII regions. Considering the small sample size used in this work—particularly the inclusion of only two young \hii\, regions, which also have large uncertainties in their expansion velocities—further confirmation of this interpretation will require higher-resolution 2D spectroscopy to resolve blended kinematic components along the line of sight for more accurate estimation of expansion velocities, along with an expanded sample that specifically includes more young \hii\, regions. 
}

\authorrunning{Lin Ma et al.}            
\titlerunning{Size-$\sigma$ relation in Galactic \hii\, regions}  
\maketitle

%
\section{Introduction}           
\label{sect:intro}

The dynamical structure of interstellar gas exhibits fundamental scaling relations that encode the complex physics governing the interstellar medium (ISM). The seminal work by \cite{Larson-1981} established the size-velocity dispersion relation ($\sigma\,\propto\,R^{0.38}$) for molecular clouds, providing crucial insights into ISM dynamics, where $\sigma$ is the velocity dispersion obtained by the spectral linewidth, and R is the radius of the molecular cloud. This relation, initially interpreted as evidence for supersonic turbulence, has been extensively studied with reported slopes varying from 0.2 to 0.7 depending on environment and scale \citep[e.g.][]{Solomon-1987, Caselli-1995,  Heyer-2004, Bolatto-2008, Zhou-2022, Xing-2022, Feng-2024}.  Such variations likely reflect different dominant physical processes, including turbulence, gravitational collapse, stellar feedback and galactic shear \citep[e.g.][]{Larson-1981, Wada-2002, Elmegreen-2004, Bolatto-2008, Zhou-2022, Xie-2025}.

This correlation has also been found in \hii\, regions. \cite{Melnick-1977} discovered a strong size-$\sigma$ relation in the \ha\, emission from giant \hii\, regions in late-type spiral galaxies, finding a scaling of $\sigma \propto R^{0.44}$. 
\cite{Terlevich-1981} further demonstrated that the size-$\sigma$ relation for giant \hii\, regions closely resembles those of elliptical galaxies and globular clusters. Combining these three types of systems, they derived a steeper relation of $\sigma \propto R^{0.54}$. These findings suggest that giant \hii\, regions are self-gravitating systems, with their emission-line widths reflecting motions governed by the local gravitational field.
\cite{Roy-1986} studied 47 giant extragalactic \hii\, regions in 16 nearby galaxies,  $\sigma \propto R^{n}$, with n $>$ 0.63, suggesting hydrodynamic turbulence—rather than virial equilibrium—dominates their dynamics. 
\cite{Fuentes-Masip-1995} reported a strong correlation between size and $\sigma$ for \hii\, regions in NGC 4449, though they did not provide quantitative fitting results for this relationship. Their analysis was limited to regions larger than 30 pc in size.
\cite{Fuentes-Masip-2000} studied these giant \hii\, regions and found $\sigma \propto R^{n}$, with n$>0.67$, and this correlation only hold for nebulae with high \ha\, surface brightness and supersonic Gaussian profiles. The fitted exponents are consistent with virial equilibrium.
\cite{Monreal-Ibero-2007} identified 12 external star-forming regions in Ultraluminous Infrared Galaxies (ULIRGs), finding that most obey the same size-$\sigma$ relation as globular clusters and elliptical galaxies, suggesting they are stable, self-gravitating entities. 
Despite their larger sizes and higher velocities, high-redshift (z = 0-2) star-forming regions follow the similar $\sigma \propto R^{0.42}$ scaling as local \hii\, regions, suggesting they represent scaled-up versions of these local systems \citep{Wisnioski-2012}.
Similar work has been done by \citep{Swinbank-2012}, who found a steeper relation of $\sigma \propto R$. This difference may arise from the larger scatter in their sample.
\cite{Ambrocio-Cruz-2016} presented a kinematical and photometric catalog of 210 \hii\, regions in the Large Magellanic Cloud (LMC), reporting sizes typically $>20$ pc and examining their size-$\sigma$ relation. While they characterized this relationship observationally, the study did not provide a formal fitting result for the correlation.
Using high spectral and spatial resolution observations of the nearest starburst galaxy IC 10, \cite{Cosens-2022} identified 46 individual \hii\, regions with an average radius of 4.0 pc and $\sigma$ of $\sim$16 km s$^{-1}$. Their analysis revealed that none of these regions are virialized, showing clear evidence of the ongoing expansion of these regions. Notably, IC 10's \hii\, regions deviate from the established size-$\sigma$ relation observed in other galaxies. 
Besides, the data from \cite{Ambrocio-Cruz-2016} are used in the analysis, and follow the trends well. 

Previous studies have primarily focused on relatively large \hii\, regions (R $>$ 20 pc) in extragalactic systems, but whether smaller \hii\, regions exhibit significant deviations from established scaling relations remains unclear. Using data from the Large Sky Area Multi-Object Fiber Spectroscopic Telescope (LAMOST) \citep{Wang-1996, Su-Cui-2004, Cui-2012, Zhao-2012, Luo-2015}, we selected a spatially resolved sample of iosolated H II regions (R $<$ 20 pc) and analyzed their radial profiles, including optical emission-line fluxes, flux ratios, and radial velocities. We further investigated the evolution of the escape fraction during \hii\, region expansion \citep{Zhang-2025}. In this work, we use this sample to explore whether these smaller \hii\, regions follow the known size-$\sigma$ relation or exhibit distinct behaviors.

This paper is organized as follows. Section 2 describes the Galactic \hii\, region sample, and Section 3 outlines the procedures for data reduction and velocity dispersion measurement. The results are given in Section 4, followed by a broader discussion in Section 5. Finally, the conclusions are summarized in the last section.

\section{Sample construction}
\label{sect:sample}

In this study, we use the \hii\, region sample from \cite{Zhang-2025}. We briefly describe the observational data and sample selection criteria, as well as the previously established parameters for the \hii\, regions.

The most complete census of Galactic \hii\, regions to date was established by \cite{Anderson-2014} using mid-infrared morphology data from the Wide-field Infrared Survey Explorer (WISE) satellite \citep{Wright-2010}. This sample improves on previous efforts by resolving \hii\, region complexes into multiple sources and by removing duplicate sources and contaminants such as planetary nebulae and supernova remnants. We use Version 2.2 of this catalog (publicly available at http://astro.phys.wvu.edu/wise/), which contains 8412 sources in total, including 2210 classified as ``Known" \hii\, regions through detections of radio recombination lines or \ha\, emission.  For each region, we adopt the catalog's circular radius (enclosing the MIR emission) as the \hii\, region size (R).

Launched in 2018, LAMOST Medium-Resolution Spectroscopic Survey (LAMOST MRS) has been providing simultaneous 4950-5350 \AA\, (blue) and 6300-6800 \AA\, (red) spectra with a resolution of $\sim$7500 ($\lambda/\Delta\lambda$), enabling $\sim$1 $\rm km\,s^{-1}$ radial velocity measurements for diverse studies including binarity, stellar pulsation, star formation, emission nebulae, Galactic archaeology, exoplanet host stars, and open clusters \citep{Liu-2020, ZhangB-2021}. The LAMOST Medium-Resolution Survey of Galactic Nebulae (MRS-N), focuses on emission nebulae in the Galactic plane (40\degr\ $<$ l $<$ 215\degr, $|b| < $ 5\degr), covering approximately 1700 deg$^2$. This survey provides key optical emission lines including \ha, \nii $\lambda\lambda$6548,6584, \sii $\lambda\lambda$6717,6731, and \oiii $\lambda\lambda$4959,5007 \citep{Wu-2021}. The wavelength calibration, corrected using sky lines, achieves precision with systematic RV deviations of 0.2-0.5 $\rm km\,s^{-1}$ \citep{Ren-2021}. Geocoronal \ha\, contamination is removed through a correlation analysis between the H$_{\alpha,\rm sky}$/OH $\lambda$6554 ratio and solar altitude \citep{ZhangW-2021}. The complete data reduction pipeline and products are summarized in \cite{Wu-2022}.

We cross-matched the LAMOST MRS-N data with the WISE \hii\, region catalog, selecting only ``Known \hii\, regions" that are isolated from bright neighbors within 2R and have angular sizes exceeding 3$\farcm$5  to ensure adequate fiber coverage. Our final sample consists of 10 regions, with ID from \hii\,-01 to \hii\,-10, their WISE name, central coordinates, and R values provided in Table \ref{tbl:pars}. 

\cite{Zhang-2025} identifies OB stars or young stellar objects (YSOs) within 1R of each \hii\, region, sourcing OB stars from \citet{Liu-2019,Xu-2021} or SIMBAD and YSOs from SIMBAD. 
We employ the photogeometric parallax distances from \citet{Bailer-Jones-2021} to estimate distances to these H II regions. This method provides higher accuracy than geometric distances, especially for stars with less reliable parallax measurements.
For regions with a single OB star or YSO, its parallax distance is adopted; for multiple sources, a weighted distance is computed using $\theta^{-2}$ as a weighting factor based on angular distance ($\theta$) to the \hii\, region center. This method yielded distances ranging from 0.7 to 5.3 kpc for all ten \hii\, regions. The WISE angular radii were then converted to physical radii using the derived distances, yielding R values ranging from 2.4 to 16.3 pc. 

For each \hii\, region, \cite{Zhang-2025} analyzed the 1D radial profile of \ha\, velocity and performed polynomial fitting. The expansion velocity (V$_{exp}$) was derived as the difference between the radial velocities at r = 0.2R and  r = R.

The classification as ionization-bounded stems from the close agreement between the optical radius (r$_{\rm H\alpha}$) derived from the \ha\, radial profile and R. In contrast, the other eight HII regions exhibit r$_{\rm H\alpha}$ values significantly larger than R, leading to their classification as matter-bounded HII regions.

The kinematic ages were derived from the physical sizes and expansion velocities of the regions. HII-03 and HII-04 yield ages of $\sim$ 0.2 Myr and 0.3 Myr, respectively, indicating that they are young HII regions. The other eight regions, with ages ranging from 0.6 to 12 Myr, are therefore classified as older \hii\, regions in this study.

The classifications, derived distances, physical radii, and expansion velocities are presented in Table \ref{tbl:pars}.

\begin{table}
    \centering 
    \begin{adjustbox}{angle=90} 
        \begin{minipage}{25cm}  
            \begin{adjustbox}{max width=\linewidth} 
                \begin{tabular}{@{}lccccccccccccccc@{}}
                \noalign{\smallskip}\hline\hline
                (1)    & (2)              & (3)      & (4)     & (5)             & (6)               & (7)                 & (8)                 & (9)                & (10)                     & (11)               & (12)             &  (13)               & (14)              & (15)                 \\
                ID     & WISE Name        & type     & R   & dist            & R            & $V_{exp}$           & $T_e$ (S1-1)        & $\sigma$ (S1-1)    & $T_e$ (S1-2)             & $\sigma$ (S1-2)    & $T_e$ (S2-1)     &  $\sigma$ (S2-1)    & $T_e$ (S2-2)      & $\sigma$  (S2-2)     \\
                &      &                  &           (arcsec) & (pc)            & (pc)              & ($\rm km\,s^{-1}$)  & (K)                 & ($\rm km\,s^{-1}$) & (K)                      & ($\rm km\,s^{-1}$) & (K)              &  ($\rm km\,s^{-1}$) & (K)               & ($\rm km\,s^{-1}$)   \\
                \noalign{\smallskip}\hline
                HII-01 & G096.345-00.157  & m-bounded & 566    & 5221 $\pm$ 244  & 14.33 $\pm$ 0.67  & 15.0 $\pm$ 5.0      &  10031 $\pm$  5664  & 23.59  $\pm$ 0.73  &  24549 $\pm$ 2267        &  18.77 $\pm$ 0.29  & 8070  $\pm$ 4820 &  20.18 $\pm$ 0.51   & 15375 $\pm$ 1513  & 11.91 $\pm$ 0.22    \\
                HII-02 & G112.212+00.229  & m-bounded & 542    & 2704 $\pm$ 117  &  7.10 $\pm$ 0.31  & 1.3  $\pm$ 2.5      &  10833 $\pm$   362  & 6.44   $\pm$ 0.10  &  11118 $\pm$  352        &   6.39 $\pm$ 0.10  & 9123  $\pm$  212 &   6.50 $\pm$ 0.04   &  9350 $\pm$  203  &  6.42 $\pm$ 0.04    \\
                HII-03 & G113.900-01.613  & i-bounded & 327    & 2712 $\pm$  73  &  4.30 $\pm$ 0.12  & 14.6 $\pm$ 9.4      &  55206 $\pm$  4962  & 7.30   $\pm$ 0.83  &  42338 $\pm$ 1729        &   7.50 $\pm$ 0.30  & 9210  $\pm$  553 &   2.63 $\pm$ 0.16   & 10667 $\pm$  472  &  1.21 $\pm$ 0.15    \\
                HII-04 & G117.639+02.275  & i-bounded & 796    & 2316 $\pm$ 140  &  8.94 $\pm$ 0.54  & 15.1 $\pm$ 8.9      &  11340 $\pm$  1042  & 7.46   $\pm$ 0.29  &  12720 $\pm$  808        &   7.41 $\pm$ 0.22  & 7681  $\pm$  187 &   5.30 $\pm$ 0.04   &  7433 $\pm$  170  &  5.30 $\pm$ 0.03    \\
                HII-05 & G136.448+02.519  & m-bounded & 317    & 5345 $\pm$ 704  &  8.21 $\pm$ 1.08  & 3.0  $\pm$ 4.6      &  18666 $\pm$   684  & 8.21   $\pm$ 0.14  &  15849 $\pm$  525        &   8.55 $\pm$ 0.09  & 10390 $\pm$  490 &   6.12 $\pm$ 0.03   & 11477 $\pm$  476  &  5.44 $\pm$ 0.11    \\
                HII-06 & G149.738-00.207  & m-bounded & 213    & 4701 $\pm$ 467  &  4.85 $\pm$ 0.48  & 0.2  $\pm$ 2.5      &  11266 $\pm$  1436  & 7.75   $\pm$ 0.35  &  11377 $\pm$ 1445        &   8.04 $\pm$ 0.35  & 14581 $\pm$ 1713 &   5.72 $\pm$ 0.47   & 15220 $\pm$ 1742  &  5.83 $\pm$ 0.47    \\
                HII-07 & G168.750+00.873  & m-bounded & 1248   & 1856 $\pm$  58  & 11.23 $\pm$ 0.35  & 2.3  $\pm$ 7.3      &  --                 & 8.41   $\pm$ 0.22  &   5055 $\pm$ 1004        &  10.55 $\pm$ 0.18  & --               &   6.80 $\pm$ 0.20   &  --          &  9.37 $\pm$ 0.15    \\
                HII-08 & G173.588-01.606  & m-bounded & 1298   & 2591 $\pm$  74  & 16.31 $\pm$ 0.46  & 2.4  $\pm$ 2.7      &  6327  $\pm$   291  & 10.57  $\pm$ 0.05  &   7916 $\pm$ 185         &   9.30 $\pm$ 0.03  & 6656  $\pm$  201 &  10.18 $\pm$ 0.03   &  7718 $\pm$  189  &  9.76 $\pm$ 0.03    \\   
                HII-09 & G201.535+01.597  & m-bounded & 790    & 3565 $\pm$ 145  & 13.65 $\pm$ 0.56  & 8.3  $\pm$ 8.3      &  19532 $\pm$   553  & 9.75   $\pm$ 0.08  &  13892 $\pm$ 294         &   7.79 $\pm$ 0.05  & 7334  $\pm$  426 &  12.70 $\pm$ 0.04   & 10210 $\pm$  213  &  9.17 $\pm$ 0.03    \\
                HII-10 & G202.968+02.083  & m-bounded & 730    & 681  $\pm$  14  &  2.41 $\pm$ 0.05  & 0.3  $\pm$ 10.0     &  11155 $\pm$   136  & 5.06   $\pm$ 0.03  &  11054 $\pm$ 135         &   4.91 $\pm$ 0.03  & 10502 $\pm$  127 &   4.47 $\pm$ 0.03   & 10535 $\pm$  126  &  4.22 $\pm$ 0.03    \\
                \noalign{\smallskip}\hline
                \end{tabular}
            \end{adjustbox}
            \caption{Parameters of the ten observed \hii\, regions. Col. 1 shows the region IDs (HII-01 to HII-10). Col. 2 gives the WISE source names from \cite{Anderson-2014}. Col. 3 specifies the H II region boundary type: matter-bounded or ionization-bounded. Cols. 4-6 include the angular radius (in arcsec), distance (in pc), and physical size (in pc). Col. 7 lists the expansion velocity of the HII shell. Cols. 8 and 9 present the electron temperature (derived from the H$\alpha$ and \nii\, line widths) and the final velocity dispersion (corrected for instrumental, thermal, and fine-structure broadening), respectively, both based on method S1-1. The corresponding results from methods S1-2, S2-1, and S2-2 are given in Cols. 10/11, 12/13, and 14/15, respectively. }
            \label{tbl:pars}
        \end{minipage}
    \end{adjustbox}
\end{table}

\begin{figure}[htbp]
\centering
\includegraphics[width=\textwidth]{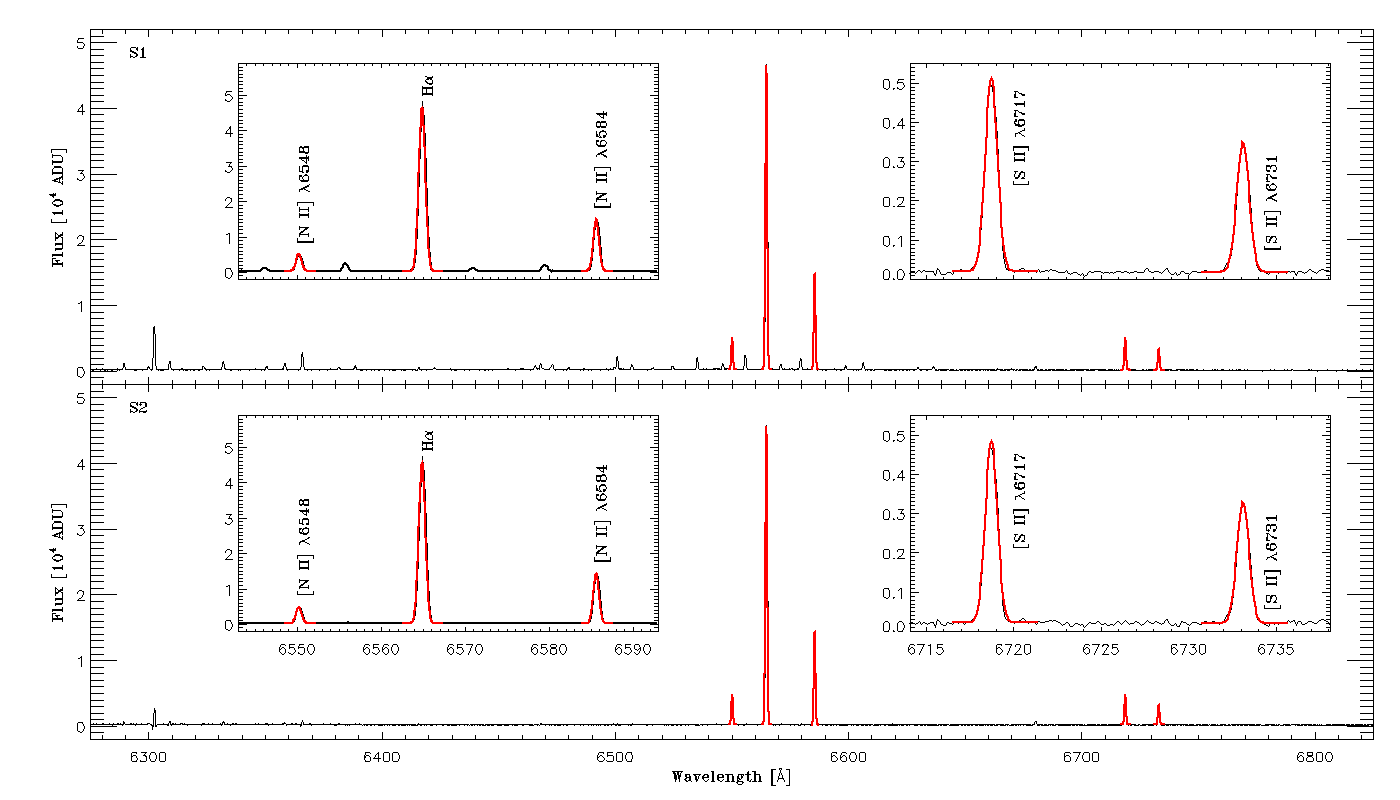}
\caption{A representative spectrum from LAMOST MRS-N. The upper panel displays the spectrum reduced by subtracting the geocoronal H$\alpha$ emission (method S1), while the lower panel shows the result of subtracting all the skylines and the foreground/background emission from DIG (method S2). Insets provide a detailed view of the nebular emission lines, including H$\alpha$, \nii\, and \sii\,. The results of a single Gaussian fit are overplotted in red. As the spectrum is not flux-calibrated, the flux is given in Analog-to-Digital Units (ADU).}
\label{fig:example_spec}
\end{figure}

\section{data reduction and parameter measurement}
\label{sect:pars}

\subsection{Subtraction of diffuse ionized gas}

As described in \S\, \ref{sect:sample}, the spectra have been processed with subtraction of geocoronal \ha\, emission \citep[see details in ][]{ZhangW-2021}; we hereafter refer to this method as S1. To accurately measure emission line widths, and considering that the spectra of \hii\, regions are often contaminated by foreground/background emission from diffuse ionized gas (DIG), we developed an additional method to subtract this DIG component. This approach, designated S2,  proceeds as follows. For each LAMOST plate, we first exclude fibers associated with \hii\, regions and SNRs. We then select the 20 fibers with the weakest \nii\,$\lambda$6584 emission to represent the DIG. These individual spectra are scaled using the intensity of the OH $\lambda$6554 sky line as a reference before being combined into a final, median DIG spectrum. The median DIG spectrum is scaled to match the intensity of the OH $\lambda$6554 sky line in each individual \hii\, region spectrum and is then subtracted from it.

We compare the spectra reduced by methods S1 and S2 in the upper and lower panels of Figure \ref{fig:example_spec}, respectively. The flux of the nebular lines from S2 is smaller than that from S1. Method S1 leaves numerous sky emission lines (except the geocoronal H$\alpha$ line) in the spectrum, whereas method S2 simultaneously removes all skylines and DIG emissions. Furthermore, the nebular [O I]$\lambda$6300 line is clearly resolved in the S2 spectrum, while it is blended with the sky component at a similar wavelength in the S1 spectrum. However, method S2 is only applicable in regions with strong nebular emission, such as \hii\, regions, SNRs, PNe, or stars contaminated by DIG. For studies targeting the DIG itself \citep[eg.][]{Wen-2025}, data products from S1 remain the preferred choice.

\subsection{Integrated spectra}

Since the LAMOST MRS-N data have not been flux-calibrated and the fiber efficiency varies across the focal plane, we perform relative flux calibration for each fiber using the skyline OH $\lambda$6554. For a given plate containing N fibers, the calibration is applied to the $i$-th fiber as follows:
\begin{equation}
F_i({\rm H}\alpha) = \frac{F_i^o({\rm H}\alpha)}{F_i^o(6554)}\frac{\sum\limits_{i=1}^{N} F_i^o(6554)}{N}
\end{equation}
where  $F_i^o({\rm H}\alpha)$ and $F_i^o(6554)$ represent the observed fluxes of \ha\, and OH $\lambda$6554 from the $i$-th fiber, respectively.

\begin{figure*}
    \centering
    \includegraphics[width=\textwidth]{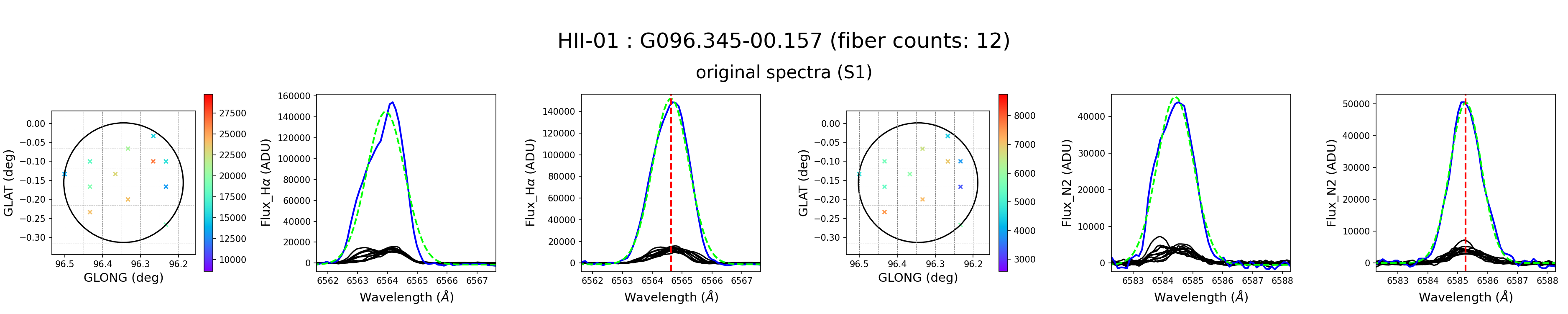}
    \includegraphics[width=\textwidth]{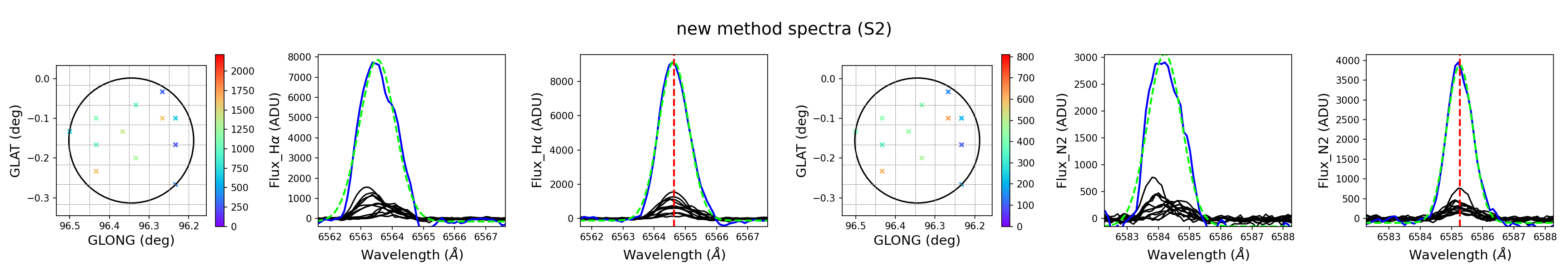}
    \includegraphics[width=\textwidth]{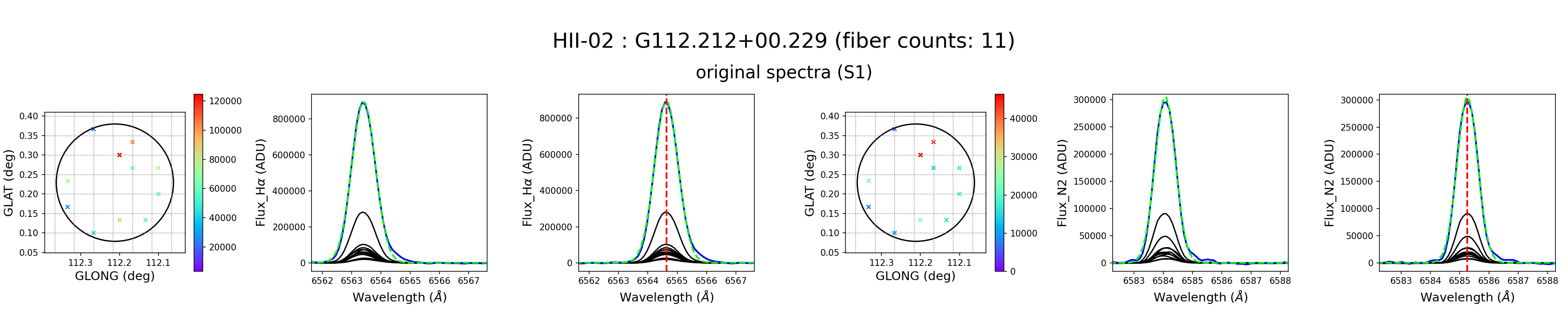}
    \includegraphics[width=\textwidth]{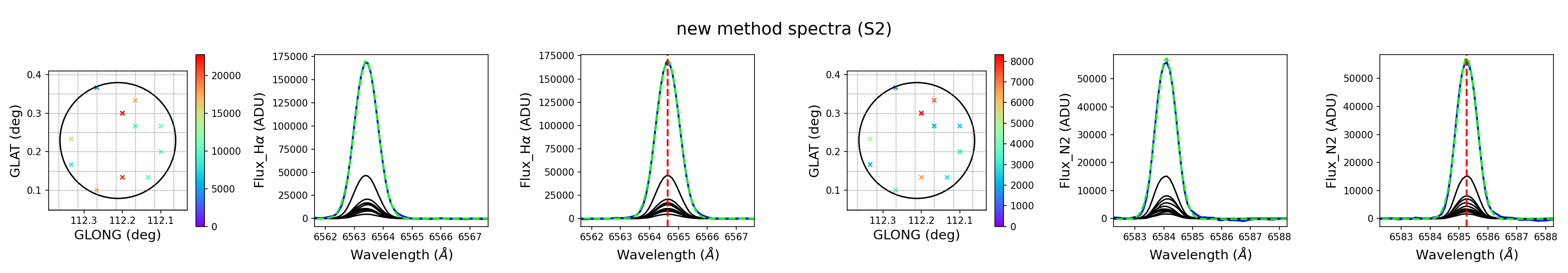}
    \caption{Fiber positions and integrated spectra. The data for each \hii\, region (labeled with its ID, WISE name, and fiber count within 1R region) is displayed across two rows. The first row presents the result of the original reduction method, which only subtracts the geocoronal \ha\ emission. The second row shows the result of the second method, which additionally subtracts the foreground and background diffuse emissions. Each row contains six panels: the first shows the fiber positions (crosses) color-coded by \ha\ flux, overlaid with $3'\times3'$ grids; the second and third panels present the observed and velocity-shifted \ha\ spectra, respectively, both displaying individual fibers (black), the co-added integrated spectrum (blue), and the best-fit Gaussian (green), with the latter also marking the \ha\ vacuum wavelength centroid (red dashed line); panels four to six follow the same format for the corresponding \nii\ data.}
    \label{fig:HII_spectra_shifted}
\end{figure*}

\begin{figure}[htbp]
    \centering
    \includegraphics[width=\textwidth]{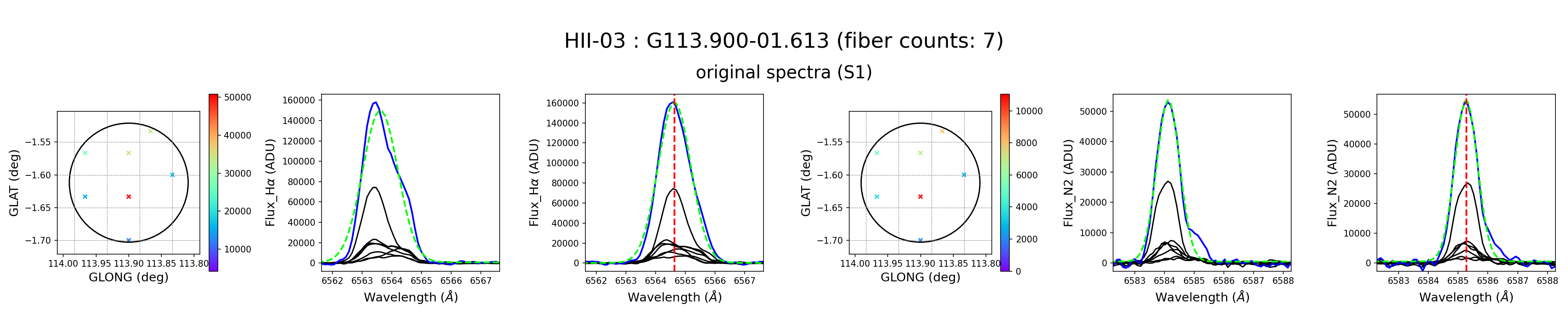}
    \includegraphics[width=\textwidth]{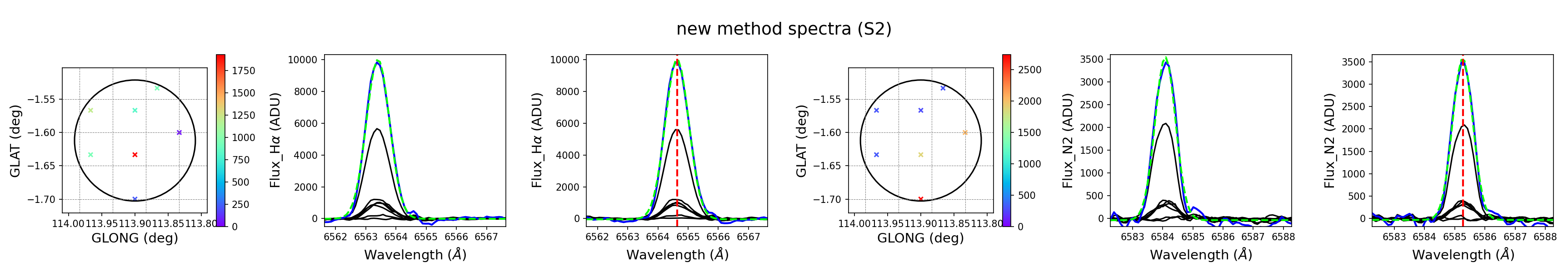}
    \includegraphics[width=\textwidth]{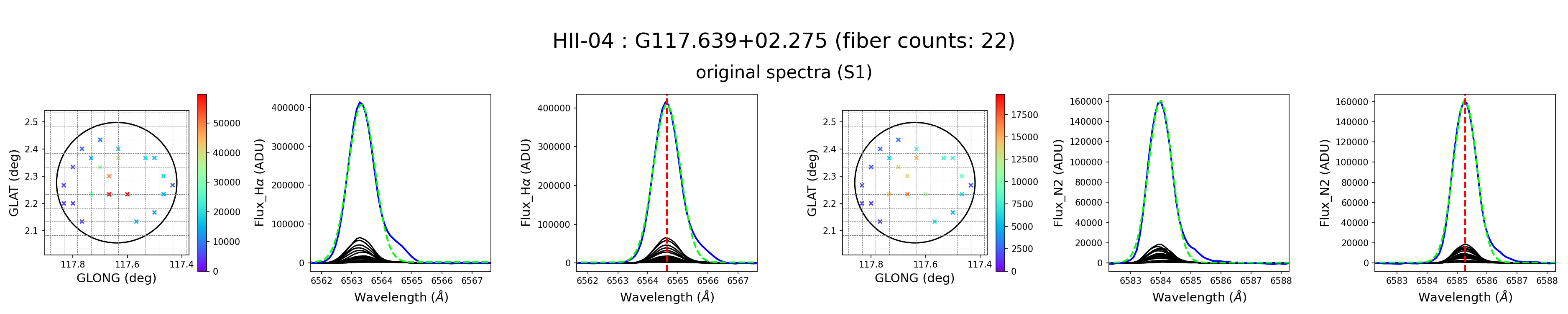}
    \includegraphics[width=\textwidth]{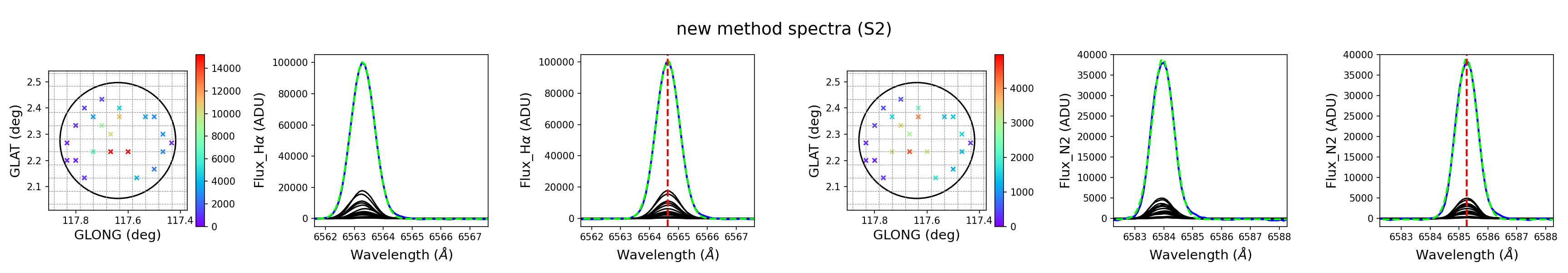}
    \includegraphics[width=\textwidth]{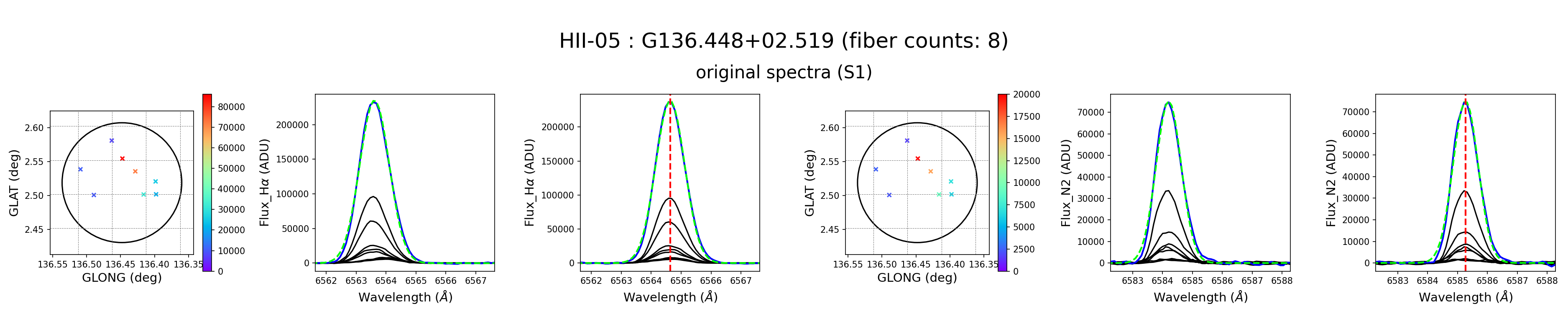}
    \includegraphics[width=\textwidth]{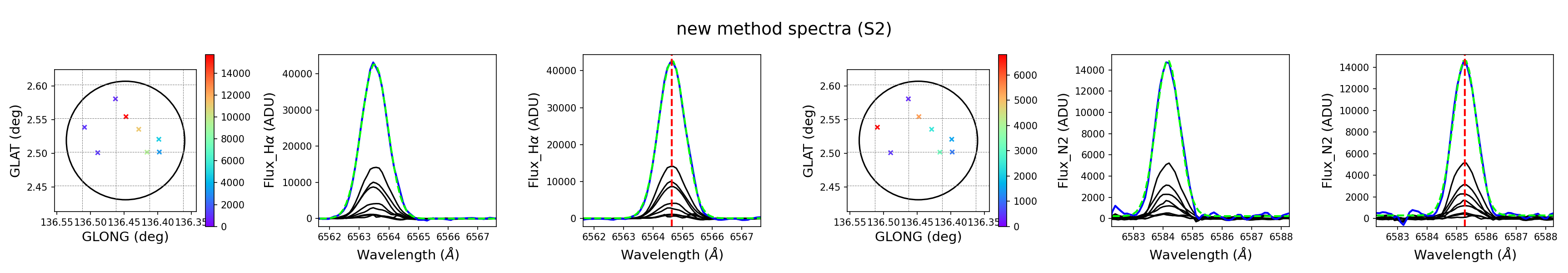}
    \includegraphics[width=\textwidth]{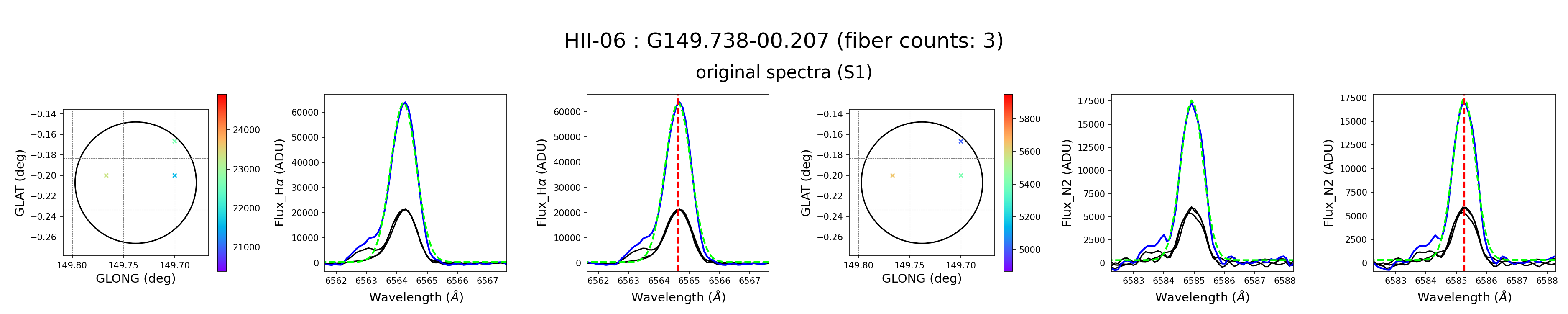}
    \includegraphics[width=\textwidth]{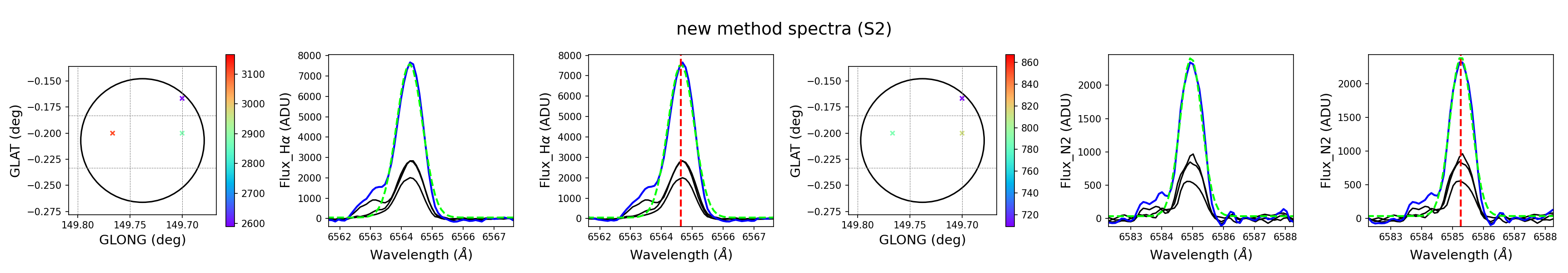}
    \continuedcaption{} 
\end{figure}

As the field of view of LAMOST is circular, adjacent plates must overlap to ensure full coverage of a continuous sky region. While fiber assignment in these overlap regions generally changes between different plates (except for a few high-priority stars), these overlap regions do exhibit a higher sampling rate than areas covered by only a single plate. It is therefore essential to resample the data to a uniform grid before combining the spectra from each  \hii\, region. Given a median seperation  of $\sim 3\arcmin$ between adjacent fibers \citep{Wen-2025}, we resample the data into $3\arcmin \times 3\arcmin$ bins. For each 2D bin, when only one fiber is present we retain it, while for bins containing multiple fibers we select the one with the highest signal-to-noise ratio (S/N) of \ha\, emission. For each fiber, we obtain the total E(B-V) between the target and the observer using a 3D reddening map \citep{Green-2019}. For simplicity, we adopt a single distance for each H II region, as listed in Table \ref{tbl:pars}. We then deredden the \ha\, spectrum using the extinction law from \cite{Fitzpatrick-1999} with the default value of R$_V$ = 3.1. The 2D grids and final fiber positions are shown as dashed lines and filled circles, respectively, in Figure \ref{fig:HII_spectra_shifted}, with the fibers color-coded by their flux in Analog-to-Digital Units (ADU).

\begin{figure}[htbp]
    \centering
    \includegraphics[width=\textwidth]{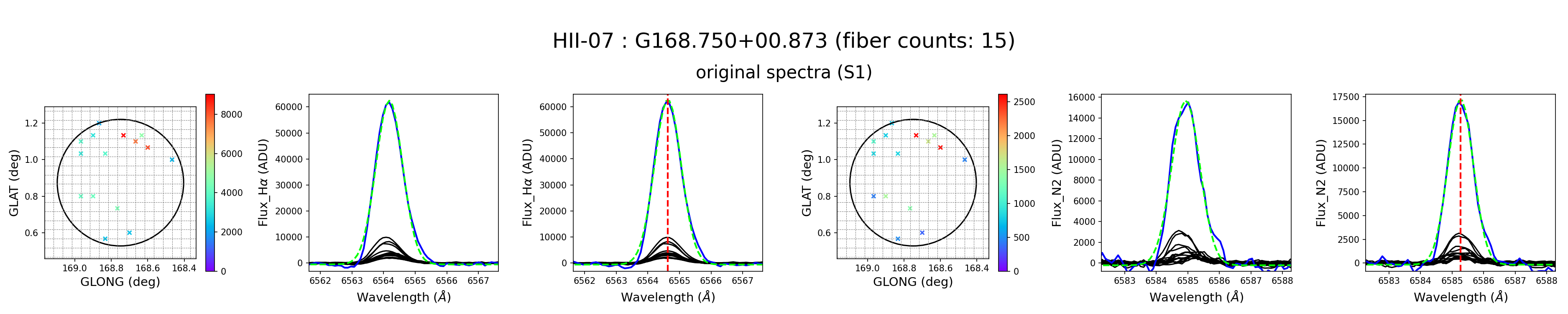}
    \includegraphics[width=\textwidth]{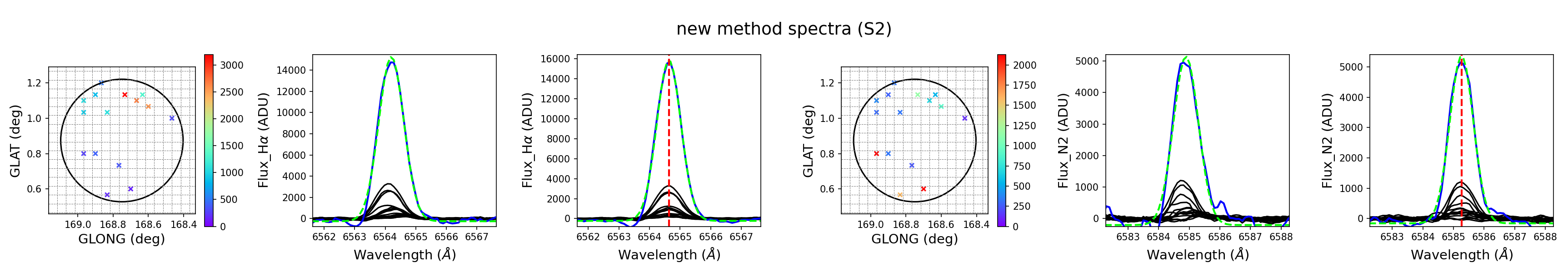}    
    \includegraphics[width=\textwidth]{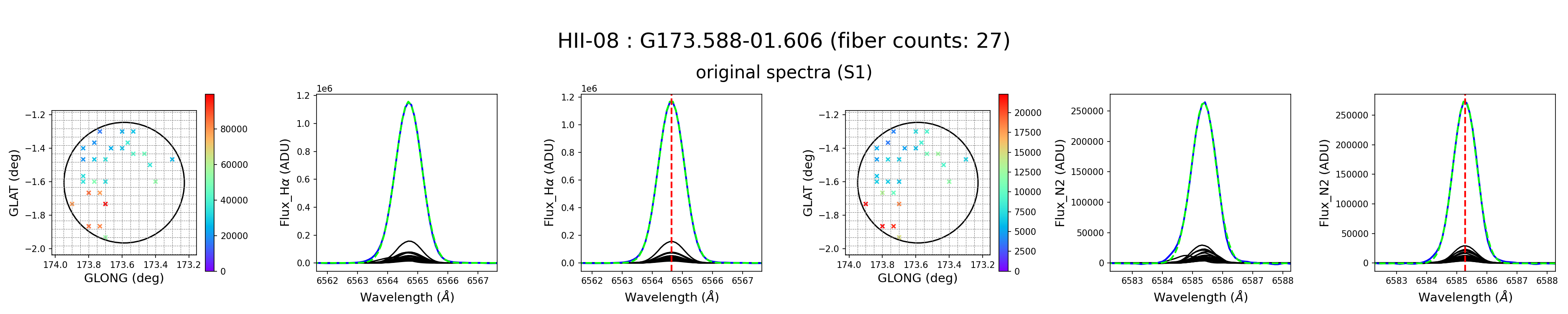} 
    \includegraphics[width=\textwidth]{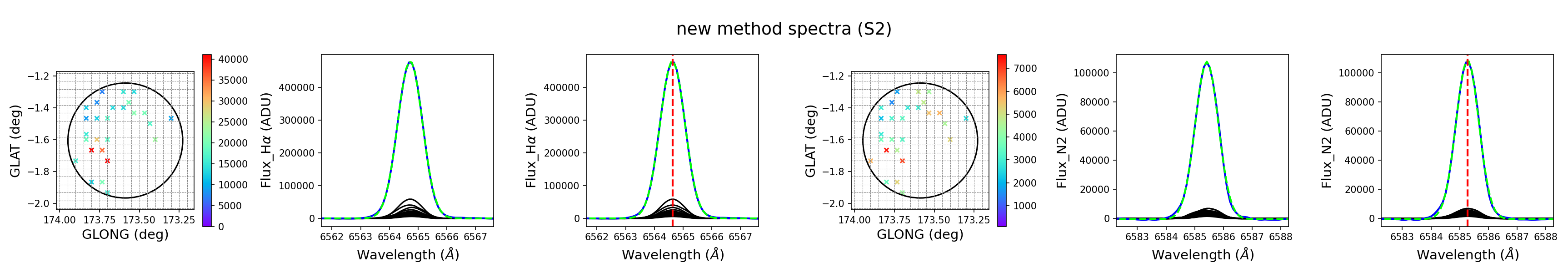}
    \includegraphics[width=\textwidth]{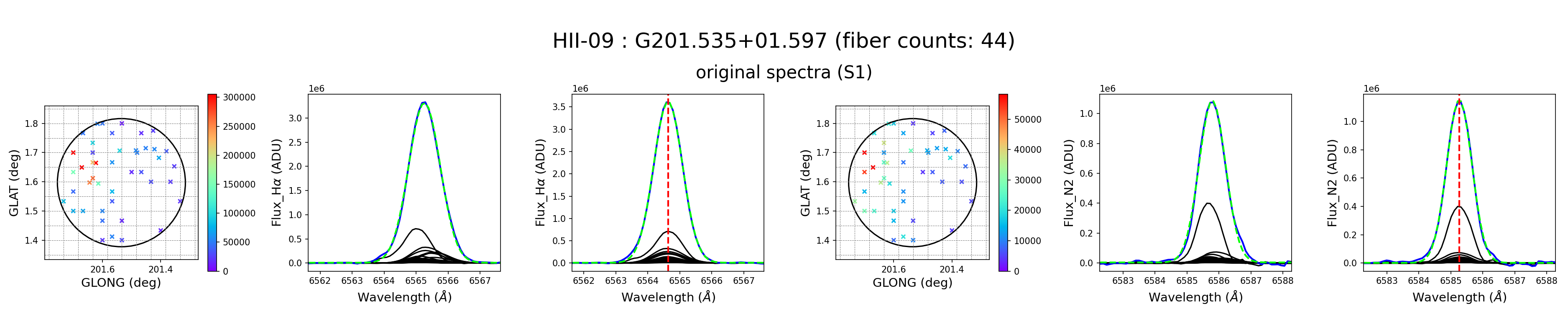}
    \includegraphics[width=\textwidth]{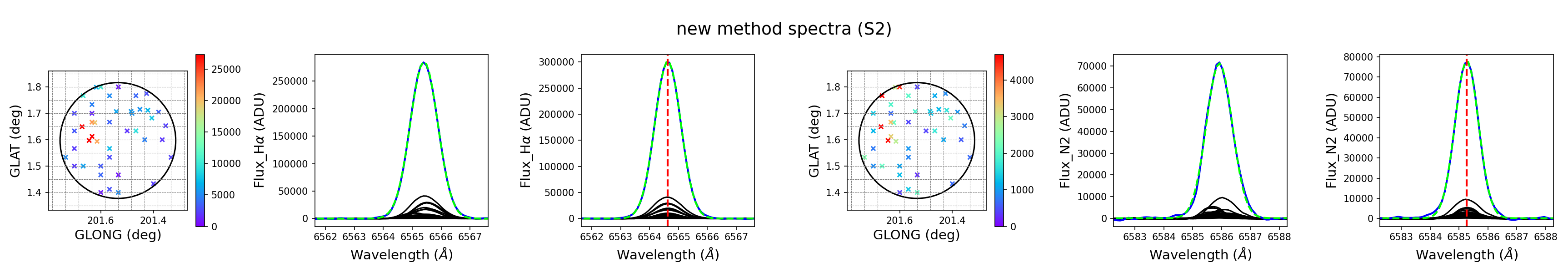}
    \includegraphics[width=\textwidth]{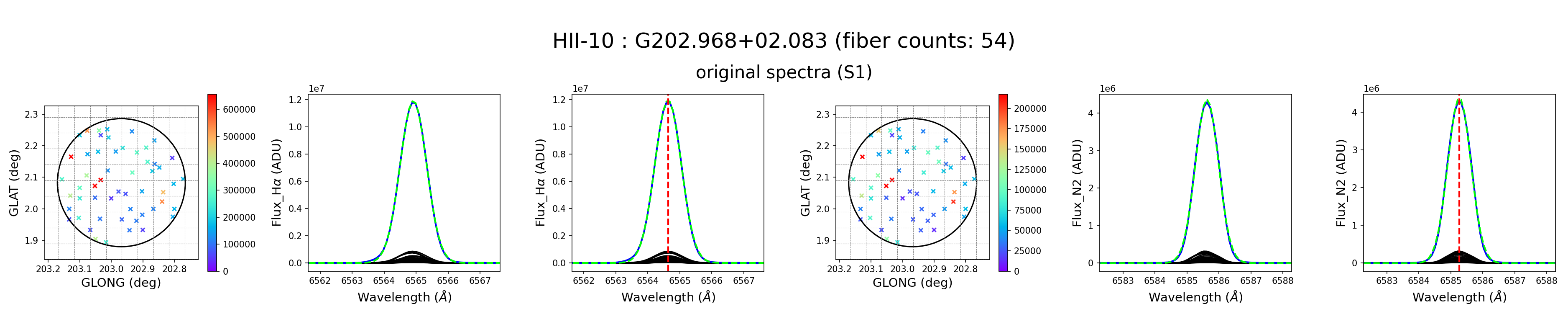} 
    \includegraphics[width=\textwidth]{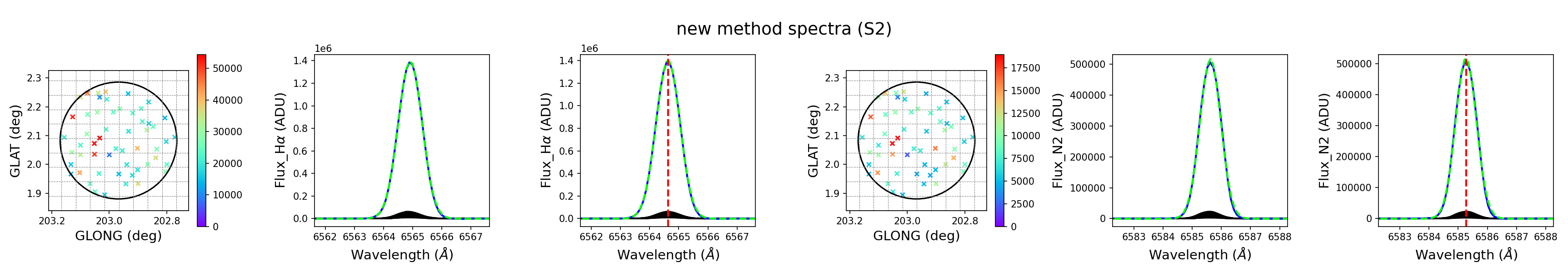}    
    \continuedcaption{}    
\end{figure}

For each \hii\, region, the variation in radial velocity among individual spectra will affect the measurement of the line width if they are coadded directly without removing the differences in their velocity centroids. In this study, to investigate how this effect influences the size-$\sigma$ relation, we retain the results from direct spectral co-addition without velocity shifting. Combining the two co-addition methods with the two sky/DIG subtraction approaches yields four distinct parameter sets for our analysis:
\begin{itemize}
    \item S1-1: The original method with geocoronal H$\alpha$ subtraction; 
    \item S1-2: The original method with geocoronal H$\alpha$ subtraction, plus removal of the differing velocity centroid; 
    \item S2-1: The new method with foreground/background subtraction, which includes all the skylines and nebular emissions from H$\alpha$, \nii, \sii\,, and [O I]; 
    \item S2-2: The new subtraction method (S2-1), plus removal of the differing velocity centroid.
\end{itemize}

In Figure \ref{fig:HII_spectra_shifted}, for each \hii\, region, we plot the \ha\, and \nii $\lambda6584$ emission lines from each fiber within 1R in black. We then stack these individual spectra to obtain the integrated \ha\, and \nii $\lambda6584$ spectra, shown in blue. A single Gaussian fit is applied to this integrated spectrum to simultaneously determine the line centroid, line width, and integrated flux. The fitted profiles are plotted as dashed green curves. 

\subsection{Instrumental and thermal broadening}

We measure the instrumental broadening ($W_{\rm inst}$) by fitting the 6554 \AA\, skylines within a 1R region of each \hii\, region using a single Gaussian profile and adopting the median value. The intrinsic line widths of \ha\, and \nii $\lambda6584$, corrected for instrumental broadening, are then calculated as: 

\begin{equation}
W_{\rm H} = [(W_{\rm H}^{\rm obs})^2-(W_{\rm inst})^2]^{1/2},\,\,\,  W_{\rm N} = [(W_{\rm N}^{\rm obs})^2-(W_{\rm inst})^2]^{1/2}
\end{equation}
where $W_{\rm H}^{\rm obs}$ and $W_{\rm N}^{\rm obs}$ represent the observed FWHM (in $\rm km\,s^{-1}$) of \ha\, and \nii\, lines, respectively, and $W_{\rm inst}$ is the instrumental broadening width (FWHM, also in  $\rm km\,s^{-1}$).

Following the method of \cite{Reynolds-1977}, we calculate the gas electron temperature ($T_{\rm e}$) for each \hii\, region as $T_{\rm e}[K] = 23.5 (W_{\rm H})^2 \left(1 - \frac{(W_{\rm N})^2}{(W_{\rm H})^2}\right)$, where both $W_{\rm H}$ and $W_{\rm N}$ are in units of $\rm km\,s^{-1}$. The derived temperatures range from $\sim$ 5000 K to 50000 K and are listed in Table \ref{tbl:pars}, with a median value of $T_e \sim$ 10000 K. We did not calculate $T_e$ for HII-07 because the width of \ha\, is smaller than that of \nii\, in methods S1-1, S2-1 and S2-2. We therefore adopt $T_e$ = 10000 K for calculating the thermal broadening for this source in these three cases. The thermal line width is calculated as: $\sigma_{\rm th} = \sqrt{\frac{kT_e}{m}}$, where $k$ is the Boltzmann constant, $T_e$ is the gas temperature in K, and $m$ is the mass of a hydrogen atom, respectively. 

\subsection{Velocity dispersion}

In addition to instrumental ($\sigma_{\rm inst}$),  thermal ($\sigma_{\rm th}$), and fine-structure ($\sigma_{\rm fs}$) broadening  must also be considered, where $\sigma=\rm FWHM/\sqrt{8ln2}$. For fine-structure broadening, we adopt $\sigma_{\rm fs} = 3.199$ $\rm km\,s^{-1}$ for H$\alpha$ from \citep{Garcia-Diaz-2008}.

The final velocity dispersion of \ha\, then is calculated as follows: 
\begin{equation}
\sigma = \left[ (\sigma_{\rm obs})^2 - (\sigma_{\rm fs})^2 - (\sigma_{\rm th})^2 - (\sigma_{\rm inst})^2 \right]^{1/2}
\end{equation}

\begin{figure} 
  \centering
  \includegraphics[width=\textwidth]{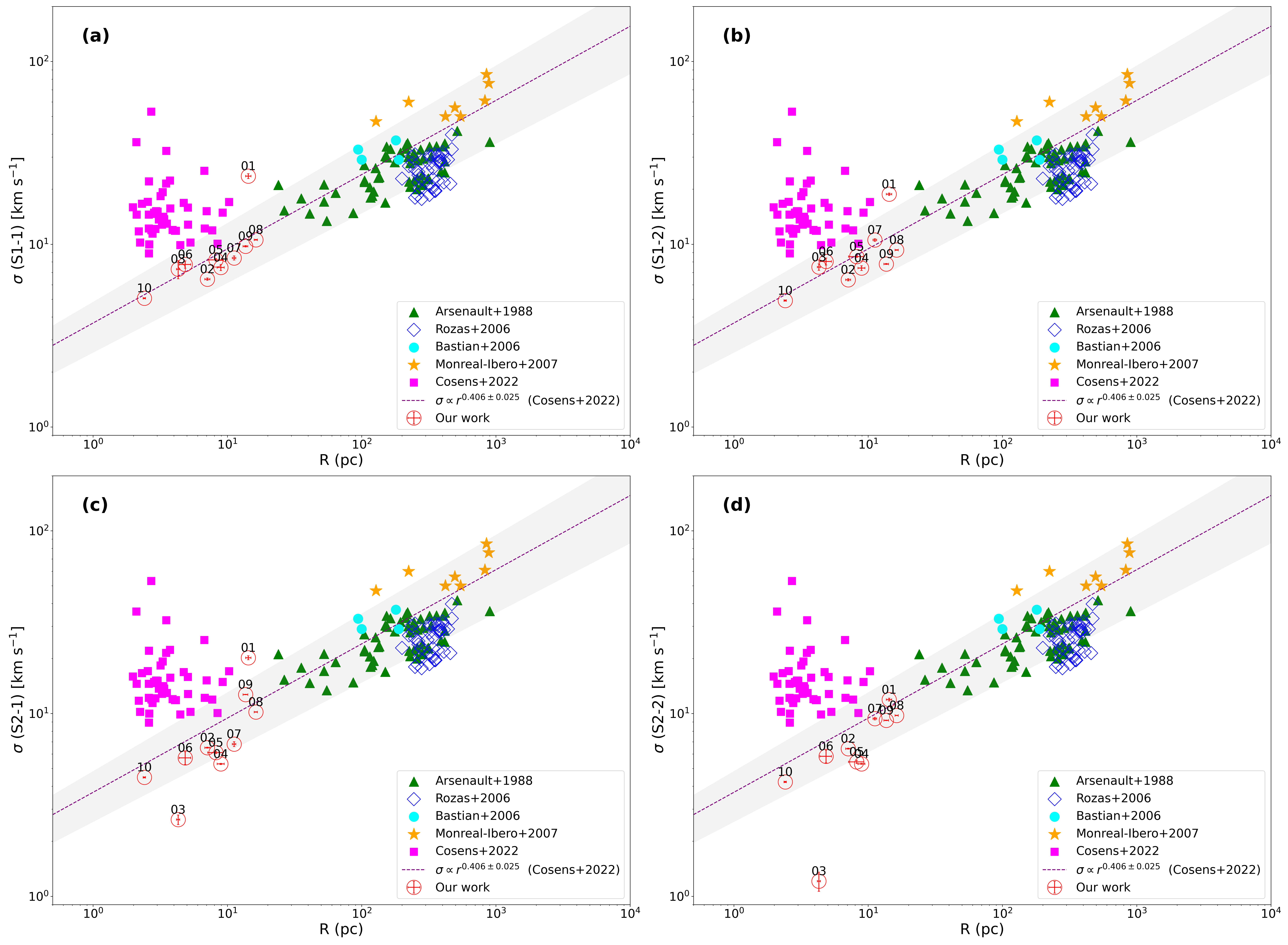}
  \caption{Size-$\sigma$ relationship of \hii\, regions. Panels (a) through (d) display the results from the S1-1, S1-2, S2-1, and S2-2 methods, respectively. The open circles represent the 10 \hii\, regions studied in this work, each labeled with its corresponding ID number, while the remaining data points are from previous studies for comparison. The dashed line shows the size-$\sigma$ relation fit from \cite{Cosens-2022}, with the shaded region indicating the uncertainties of the linear regression.} 
  \label{fig:size-sigma-relation}
  \end{figure}

  \begin{figure} 
  \centering
  \includegraphics[width=\textwidth]{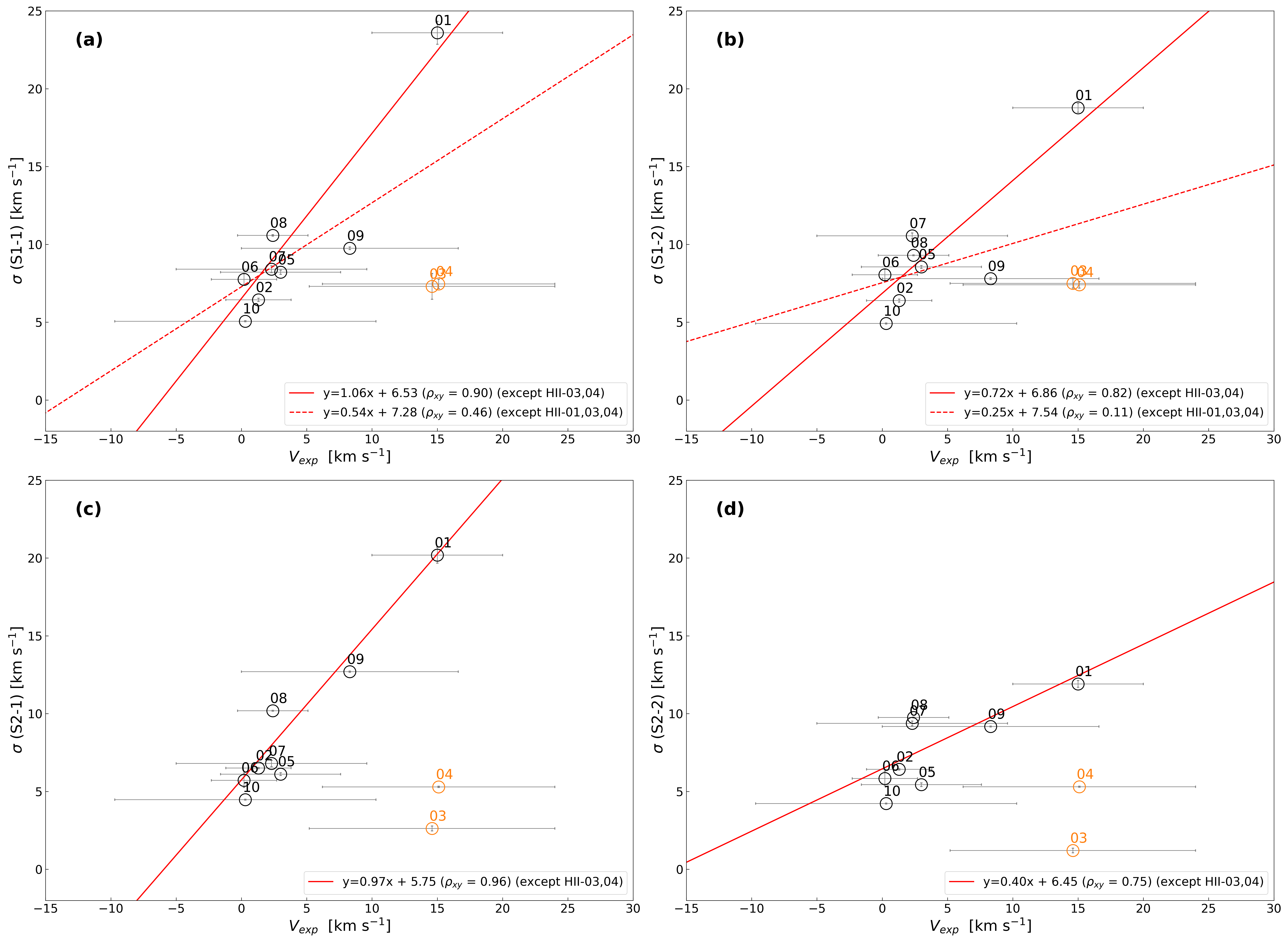}
  \caption{Correlation between the velocity dispersion and expansion velocity in \hii\, regions. Panels (a) through (d) display the results from the S1-1, S1-2, S2-1, and S2-2 methods, respectively. The ID number of each \hii\, region is displayed within its corresponding open circle. The red line shows the linear fit applied to the 8 matter-bounded \hii\, regions. In all the panels, the two ionization-bounded regions (HII-03 and HII-04) shown in orange clearly deviate from this trend and are excluded from the fit. For the linear fits shown as dashed lines in panels (a) and (b), we excluded the outlier HII-01 in addition to HII-03 and HII-04. The coefficients of these linear fits and the correlation coefficients ($\rho_{xy}$) are presented in the lower-right corners of the corresponding panels.}
  \label{fig:sigma-vexp}
  \end{figure} 

\section{Results}
\subsection{size-$\sigma$ relation}

In Figure \ref{fig:size-sigma-relation}, we investigate the size-$\sigma$ relation for \hii\, regions. Results from four distinct reduction methods are shown in separate panels for comparison. The red open circles with labeled ID numbers represent the 10 \hii\, regions analyzed in this work, while the other data points are compiled from the literature: green triangles \citep{Arsenault-1988}, blue diamonds \citep{Rozas-2006}, cyan circles \citep{Bastian-2006}, orange stars \citep{Monreal-Ibero-2007}, and magenta squares \citep{Cosens-2022}.

Rather than refitting the relation using the full sample, we adopt the slope derived by \cite{Cosens-2022}, i.e., $\sigma \propto R^n$, with n = 0.406 $\pm $ 0.025, shown as the dashed line and corresponding shaded region in light blue. Most sources in our sample fall within the shaded region, suggesting that Galactic \hii\, regions with small physical sizes ($<$ 20 pc) follow the same trend as giant extragalactic \hii\, regions.  HII-01 appears as an outlier in panels (a), (b), and (c), but it falls on the trend in panel (d). This is likely because a single-Gaussian profile inadequately describes the integrated H$\alpha$ spectrum derived from methods S1-1, S1-2, and S2-1 (shown in Figure \ref{fig:HII_spectra_shifted}), resulting in an overestimation of $\sigma$. In contrast, HII-03 is an outlier in panels (c) and (d) but follows the trend in panels (a) and (b). For this source, the line widths derived from methods S1-1 and S1-2 are also overestimated by a single-Gaussian fit; however, after subtracting the significant thermal broadening component (corresponding to $T_{\rm e}$ of 55,000 K and 42,000 K), the final $\sigma$ values fall within the normal range. Meanwhile, the spectra from methods S2-1 and S2-2 are well-fit by a single-Gaussian, and their deviation from the size-$\sigma$ relation is likely an result of the low sampling rate.

A comparison of panel (b) with (a), and (d) with (c), reveals a slight downward shift of the data points along the Y-axis after accounting for the velocity centroid difference. The small magnitude of this shift indicates that expansion effects do not dominate the velocity dispersion.

\subsection{$\sigma$-V$_{\rm exp}$ relation}

As we have measured the expansion velocity in \S\, \ref{sect:sample}, it is straightforward to investigate the possible relationship between the velocity dispersion of the integrated spectra and the expansion velocity. The results are presented in Figure \ref{fig:sigma-vexp}. Overall, a clear correlation exists for the eight matter-bounded \hii\, regions. In contrast, the two ionization-bounded \hii\, regions (HII-03 and HII-04), despite having high expansion velocities ($\sim$ 15 $\rm km\,s^{-1}$), exhibit relatively low velocity dispersions ($\lesssim$ 5 $\rm km,s^{-1}$) and thus deviate significantly from this correlation. Consequently, we excluded them from the subsequent linear fitting. The red solid lines show the linear fits to the eight matter-bounded \hii\, regions.

The near-unity slopes for methods S1-1 and S2-1 suggest that expansion contributes to the $\sigma$ derived from the integrated spectrum, though not dominantly as previously discussed. This is further supported by the shallower slopes of S1-2 and S2-2, where removal of the velocity centroid difference diminishes the expansion effect. The persistence of this effect is due, at least in part, to the fact that the line-of-sight expansion in individual spectra remains unresolved at the current spectral resolution.

HII-01 is an outlier in the size-$\sigma$ relation in the panels (a) and (b) in Figure \ref{fig:size-sigma-relation}, and also exhibits deviations in panels (a) and (b) of Figure \ref{fig:sigma-vexp},  we performed an additional fit by excluding this source, with the result shown as a red dashed line in Figure \ref{fig:sigma-vexp}. This yields a shallower slope, a change partly influenced by the position of HII-09. The shallower slope in panel (b) compared to panel (a) is consistent with the conclusion that expansion effect contributes partially to the $\sigma$ measurement. We did not perform this additional fitting in panels (c) and (d) in Figure \ref{fig:sigma-vexp}, as HII-01 aligns well with the $\sigma$-V$_{\rm exp}$ relation.

As the value of $\sigma$ depends on the analysis method, the correlation coefficient ($\rho_{xy}$) for the $\sigma$–V${\rm exp}$ relation varies across the four methods, with values of 0.90, 0.82, 0.96, and 0.75 for S1-1, S1-2, S2-1, and S2-2, respectively. Although S2-2 shows the weakest correlation among the four methods—though still strong in absolute terms—we nonetheless suggest that the $\sigma$–V$_{\rm exp}$ relation derived from S2-2 is the most physically trustworthy. This conclusion is based on its key methodological strength: the S2-2 calculation of $\sigma$ explicitly excludes contamination from foreground/background DIG and partially corrects for the effects of expansion.

\section{Discussion}

The \hii\, region sample from \cite{Cosens-2022} spans a similar range of physical sizes (R $<$ 20 pc) as our sample, yet exhibits systematically higher velocity dispersion at comparable sizes. Consequently, their sample shows significant deviation from the size-$\sigma$ relation. The authors attribute this to non-virialized dynamics, as evidenced by their analysis of the luminosity-$\sigma$ relation, which revealed an additional kinematic component beyond gravitational motion. Using a simple rotating \hii\, region model, they estimated and subsequently removed this rotational contribution from the velocity dispersion. After this correction, their data showed better agreement with the size-$\sigma$ relation, though with considerable scatter. In contrast, \citep{Zhang-2025} found no evidence of rotation in the 1D radial velocity profiles of our \hii\, regions, eliminating the need to account for rotational effects. Nevertheless, the virialization state of our sample remains an open question. We plan to investigate this through analysis of the luminosity-$\sigma$ relation using a significantly larger \hii\, region sample (Zhao et al. in prep.).

Three primary mechanisms drive turbulence in \hii\, regions: (1) stellar winds inject energy via line-driven outflows from central OB stars, generating turbulent motions; (2) ionizing radiation heats the gas, and subsequent cooling drives turbulent flows through thermal instabilities, though this effect typically lasts only $\sim$1 Myr \citep{Kritsuk-2002a, Kritsuk-2002b}; and (3) ionizing radiation also triggers supersonic expansion as overpressurized gas disrupts its surroundings, producing additional turbulence \citep{Matzner-2002, Low-2004}.

The dynamical evolution of \hii\, regions reflects the competition between these mechanisms. In our sample, the correlation for older \hii\, regions ($>$0.5 Myr) suggests dominance by expansion-driven turbulence (third mechanism), while the two younger \hii\, regions showing deviations from this trend—likely remain influenced by stellar winds and radiative heating/cooling (first and second mechanisms). We hereby clarify that ``expansion-driven turbulence" is distinct from the ``expansion effect" in the measurement of $\sigma$. The expansion effect can be reduced or eliminated through high spatial and spectral resolution IFS observations, whereas expansion-driven turbulence represents an inherent, random motion of the gas.

From a data reduction perspective, the \ha\, flux in the two ionization-bounded regions drops off sharply with radius, meaning the integrated spectrum is heavily weighted toward the central region. As a result, the velocity dispersion is narrower at a given expansion velocity. Conversely, for matter-bounded \hii\, regions, the shallower \ha\, radial profile allows the outer regions to contribute more significantly to the integrated spectrum, leading to a relatively higher velocity dispersion at the same expansion velocity.
   
The tight correlation between size and $\sigma$ in Galactic \hii\, regions offers a novel approach for distance estimation. Currently, the most precise methods for nearby \hii\, regions are parallax measurements of associated masers or OB stars. Masers in massive star-forming regions enable extremely accurate distance determinations through Very Long Baseline Interferometry (VLBI) parallax measurements, though they are primarily found in dense, ultra-compact \hii\, regions. While OB star parallax provides another reliable method, only about 13\% of \hii\, regions have identifiable associated OB stars (Zhao et al. in prep.).
Kinematic distances, derived from radial velocities and Galactic rotation models, remain widely used but suffer from significant limitations. A striking example is the W3 complex, where maser parallax measurements yield precise distances of 1.95 $\pm$ 0.04 kpc \citep{Xu-2006} and 2.04 $\pm$ 0.07 kpc \citep{Hachisuka-2006}, while kinematic distances based on radial velocities range from 3.5 to 4 kpc \citep{Megeath-2008}. This significant discrepancy demonstrates that \hii\, regions can substantially deviate from the Galactic rotation curve. This method becomes particularly unreliable toward the Galactic Center and Anti-Center directions, where uncertainties exceed 50\%, compounded by the Kinematic Distance Ambiguity (KDA) problem in the inner Galaxy. Our results demonstrates that the ``universal" size-$\sigma$ relation provides an alternative distance estimation method. By combining observed velocity dispersion (yielding physical size) with angular size measurements from imaging data, we can derive reliable distance estimates independent of traditional methods.

    \begin{figure}[htbp]
    \centering
    \includegraphics[width=\linewidth]{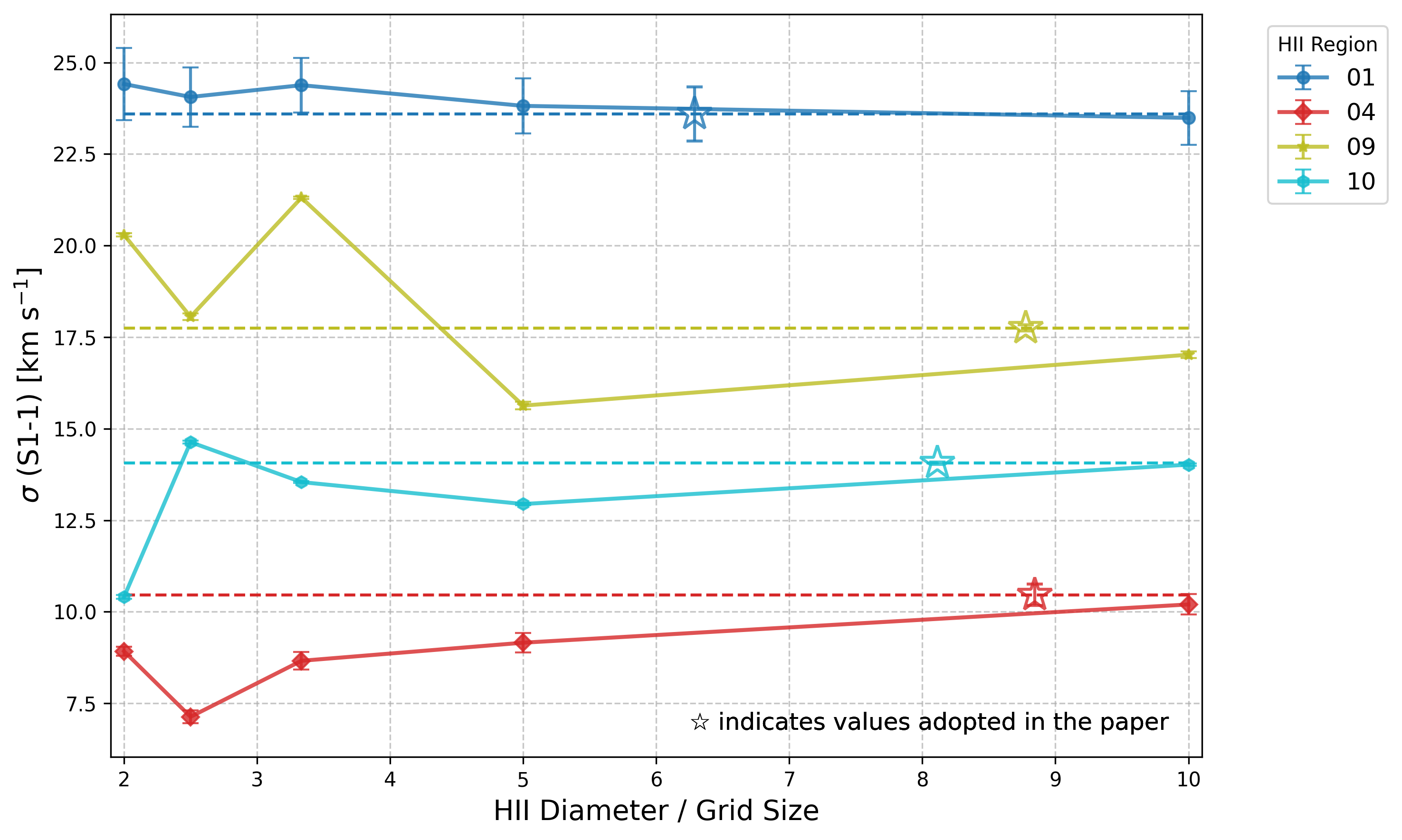}  
    \caption{Variation of $\sigma$ as a function of sampling rate, defined as the ratio of the \hii\, region diameter to the grid size. For comparison, the $\sigma$ values adopted in this study are indicated by stars and dashed lines.}
    \label{fig:sigma_gridsize} 
    \end{figure}

We acknowledge that the sampling rate can also impact the measurement of $\sigma$. To address this, we used four well-sampled \hii\, regions to calculate $\sigma$ across a range of sampling rates. The sampling rate is defined as the ratio of the \hii\, region diameter to the grid size, ranging from 2 (poor sampling) to 10 (good sampling). This range corresponds to $\sim$ 4 to 100 fibers being used in the analysis. The results are presented in Figure \ref{fig:sigma_gridsize}. For comparison, the $\sigma$ values adopted in this study (using a fixed grid size of 3$\arcmin$) are indicated by stars and dashed lines.

The deviation of the value of $\sigma$ tends to depend on the sampling rate, i.e., the lower the sampling rate, the higher the deviation from the ``true" value (e.g., as derived from IFS data). However, it is difficult to tell whether $\sigma$ will be underestimated or overestimated under poor sampling conditions, given the small sample size used in this test and the potential existence of different velocity dispersion variation patterns among different \hii\, regions. In addition, the low sampling rates for some \hii\, regions might contribute to the scatters of the size-$\sigma$ relation and $\sigma$-$V_{exp}$ relation presented in this study.

\section{Conclusion}

Based on 10 Galactic \hii\, regions selected by \citep{Zhang-2025} from the LAMOST MRS-N dataset, we measured their \ha\, velocity dispersions after removing contributions from instrumental, thermal, and natural broadening. We investigated the size-$\sigma$ relation for the sample, with the results summarized as follows:

\begin{itemize}
    \item Our sample closely follows the size-$\sigma$ relation reported by \cite{Cosens-2022} (excluding IC 10). This agreement provides strong observational evidence that small \hii\, regions (R $<$ 20 pc) exhibit the same size-$\sigma$ relation as giant \hii\, regions.  The tightness of this correlation suggests it could serve as a new distance indicator for Galactic \hii\, regions. 
    
    \item  
    The expansion of \hii\, regions contributes to the measured $\sigma$ values but is not the dominant factor. This indicates that the size-$\sigma$ relation remains applicable to spatially unresolved \hii\, regions (e.g., in high-redshift galaxies), though the $\sigma$ values are slightly overestimated due to expansion effects.
    
    \item In young, ionization-bounded \hii\ regions, the lack of a correlation between velocity dispersion and expansion velocity indicates that turbulence is primarily driven by stellar winds and ionization-induced heating/cooling processes. In contrast, the clear correlation in older, matter-bounded regions suggests that expansion-driven turbulence becomes significant.
   
\end{itemize}

While the current sample is limited—particularly as it includes only two young \hii\, regions, which also carry large uncertainties in their expansion velocities—we note that the lack of a correlation between $\sigma$ and expansion velocity suggests the need for a larger sample containing more young HII regions. In addition, because the present sample is confined to isolated HII regions, an expanded sample by including HII regions in more complex environments is also required to reassess the validity of our interpretation.
Furthermore, we plan to use the enlarged sample to investigate the luminosity-$\sigma$ relation, testing whether compact \hii\, regions (R $<$ 20 pc) deviate from virial equilibrium. Finally, since kinematic components along the same line of sight cannot be resolved with the current data, the expansion effect introduces uncertainty in interpreting the $\sigma$-V$_{\rm exp}$ relation. Therefore, the proposed mechanism requires confirmation through higher spectral resolution 2D observations.

\normalem

\begin{acknowledgements}
This work is supported by the National Natural Science Foundation of China (NSFC) (Nos. 12090041, 12090044, and 12090040) and the National Key R\&D Program of China grant (Nos. 2021YFA1600401 and 2021YFA1600400). This work is also sponsored by the Strategic Priority Research Program of the Chinese Academy of Sciences (No. XDB0550100). The Guoshoujing Telescope (the Large Sky Area Multi-Object Fiber Spectroscopic Telescope (LAMOST)) is a National Major Scientific Project built by the Chinese Academy of Sciences. Funding for the project has been provided by the National Development and Reform Commission. LAMOST is operated and managed by the National Astronomical Observatories, Chinese Academy of Sciences. 
\end{acknowledgements}
  
\bibliographystyle{raa}
\bibliography{ms}

\end{document}